\documentclass[10pt,reqno]{amsart}
\usepackage{hyperref}
\usepackage{float}
\usepackage{mathtools}
\usepackage[all]{xy}

\usepackage{tikz-cd}
\usepackage{quiver}
\usepackage{amscd,amssymb,amsmath,latexsym,bm}
\usepackage[mathcal,mathscr]{euscript}
\usepackage{lipsum}
\usepackage{amsfonts}
\usepackage{graphicx}
\usepackage{epstopdf}
\usepackage{algorithmic}
\usepackage{hyperref}
\usepackage{xcolor}
\usepackage{upgreek}
\usepackage{enumerate}
\usepackage{ulem}
\usepackage{stmaryrd}
\usepackage{gensymb}			
\usepackage[utf8]{inputenc}
\usepackage[small,nohug]{diagrams}
\diagramstyle[labelstyle=\scriptstyle]
 
\newtheorem{theorem}{Theorem}[section]

\newtheorem{remark}[theorem]{Remark}
\newtheorem{definition}[theorem]{Definition}
\newtheorem{example}[theorem]{Example}

\numberwithin{equation}{section}
\numberwithin{figure}{section}

\newcommand{\BM}{{\mathbb B}}
\newcommand{\CM}{{\mathbb C}}
\newcommand{\NM}{{\mathbb N}}

\newcommand{\RM}{{\mathbb R}}

\newcommand{\ZM}{{\mathbb Z}}

\newcommand{\FM}{{\mathbb F}}

\newcommand{\Tt}{{\mathcal T}}

\newcommand{\Nn}{{\mathcal N}}

\newcommand{\Cc}{{\mathcal C}}

\newcommand{\sgn}{{\rm sgn}}

\begin{document}
\title{Quantum versus Population Dynamics over Cayley Graphs}

\author{Emil Prodan}

\address{Department of Physics and
\\ Department of Mathematical Sciences 
\\Yeshiva University 
\\New York, NY 10016, USA \\
\href{mailto:prodan@yu.edu}{prodan@yu.edu}}

\date{\today}

\begin{abstract} 
Consider a graph whose vertices are populated by identical objects, together with an algorithm for the time-evolution of the number of objects placed at each of the vertices. The discrete dynamics of these objects can be observed and studied using simple and inexpensive laboratory settings. There are many similarities but also many differences between such population dynamics and the quantum dynamics of a particle hopping on the same graph. In this work, we show that a specific decoration of the original graph enables an exact mapping between the models of population and quantum dynamics. As such, population dynamics over graphs is yet another classical platform that can simulate quantum effects. Several examples are used to demonstrate this claim.
\end{abstract}

\thanks{This work was supported by the U.S. National Science Foundation through the grants DMR-1823800 and CMMI-2131760.}

\maketitle


\setcounter{tocdepth}{1}

\section{Introduction and Main Statements}
\label{Sec:Introduction}

Quantum dynamics has been reproduced with classical degrees of freedom for many cases of interest. For example, there are several instances where the topological dynamics predicted for electronic systems has been actually observed for the first time with classical metamaterials (see \cite{BarlasPRB2018,ChengPRL2020,ChengPRA2021,QianArxiv2022,ChengArxiv2022} for some examples). This close relation between quantum dynamics and classical dynamics of metamaterials was and continue to be beneficial to both condensed matter and materials/metamaterials communities. In recent years, another classical platform for simulating interesting dynamical features of condensed matter systems has emerged, namely, that of population dynamics over graphs \cite{TangPRX2021}. This platform is further developed here to a point where {\it every} time-evolution of a quantum Hamiltonian over a Cayley graph of a group can be observed from a classical population dynamics.

\begin{figure}[t]
\center
\includegraphics[width=0.5\textwidth]{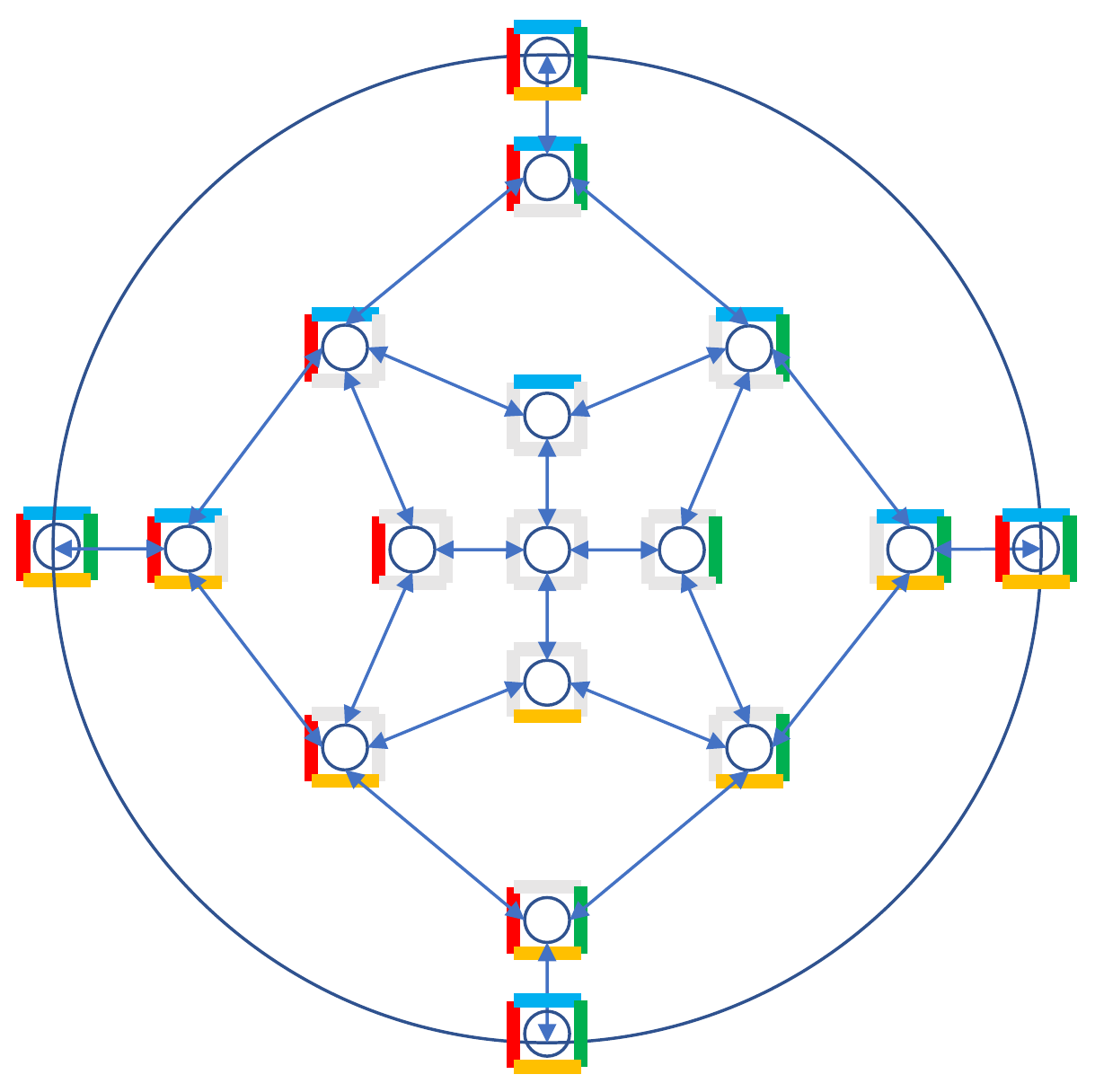}\\
  \caption{\small Abstraction of the manufacturing process of a square.
}
 \label{Fig:Manufacturing1}
\end{figure}

Population dynamics is, of course, very interesting in itself. An example of a discrete dynamics over a graph is that of population migration between different settlements. Here, the settlements are mapped into the nodes of a graph that are labeled by their geographical location, and the migration routs are mapped to the edges of the graph. Each settlement is occupied by a number of people and population dynamics is reflected in the time evolution of those numbers. There are other situations where the mapping to a graph is not so obvious, yet the perspective brought in by such mapping can be extremely valuable. Consider, for example, a product being manufactured in a factory. There are many different parts at one time inside the production line and each part evolves into another part under the manufacturing process. In this case, the problem of labeling the parts is more subtle. Indeed, although the number of distinct parts is finite, labeling them  in some order 1,2, etc., will not add much value to the abstraction of the manufacturing process. It will be much more illuminating if we use the manufacturing process itself to label the parts. An example is shown in Fig.~\ref{Fig:Manufacturing1}, where one can see all the parts involved in the assembly of a square and how they are transformed by the elementary manufacturing process. In this simple example, the elementary operations consist of attaching/removing one edge to/from an existing part (rotations of the pieces are excluded). These (reversible) elementary processes are represented by the two-headed arrows in Fig.~\ref{Fig:Manufacturing1}. The point of this example is to show that, when the parts are organized via the relations induced by the manufacturing process itself, then they naturally populate a graph whose edges represent the elementary manufacturing processes. When approached this way, various analyses of the manufacturing process can be done geometrically. For example, detecting the optimal assembly processes and how many of them exist reduces to examining the paths between the obvious points of the graph. Following this model, any assembly and self-assembly process can be rationalized using graphs and discrete dynamics over these graphs.

The different parts shown in Fig.~\ref{Fig:Manufacturing1} can be thought of as the square without the left edge, the square without top and left edges, etc.. Thus, all the nodes seen in Fig.~\ref{Fig:Manufacturing1} can be interpreted as different configurations or states of the same object, the square. This is a useful perspective because, for example, playing with a Rubik cube,  does not involve assembling, but each of the allowed elementary moves changes the cube's configuration, set by the colors of its 48 elementary squares. This example is special because, unlike for the situation described in Fig.~\ref{Fig:Manufacturing1}, an elementary move can be applied to every single configuration of the Rubik cube. In fact, an elementary move permutes the colors of the 48 squares. These elementary permutations generate a subgroup $G$ of the full permutation group of 48 squares and any configuration of the Rubik's cube is uniquely associate to an element of this $G$. Every discrete group and set of finite elements generate a Cayley graph (see subsection~\ref{Subsec:Cayley}). It turns out that playing with the Rubik cube is the same as walking on this Cayley graph. Solving the Rubik cube is equivalent to walking on a path on this graph that ends at the winning configuration of the cube. This example is not singular: If the elementary processes just permute the configurations of an object, {\it i.e.} they are bijections from the set of configurations to itself, then the information can be always organized in a Cayley graph of a subgroup of the permutation group. Thus, Cayeley graphs are interesting and occur often in this type of problems.

Dynamics over Cayley graphs can occur naturally or it can be created synthetically for different purposes. This author is more interested in the synthetic ones for the sake of visualizing various dynamical effects. We use the simple group $\ZM_N$ to exemplify. Its Cayley graph is a closed linear chain with $N$ nodes. Let us assume that the graph was populated with $N_n$ chips at position $n$, with $n$ sampling $\ZM_N$. Then we can split each of the $N_n$ chips into two stacks and move one stack at position $n-1$ and the other one at position $n+1$. The resulting algorithm is
\begin{equation}
\{N_n\}_{n \in \ZM_N} \mapsto \{N'_n = \tfrac{1}{2}(N_{n-1} + N_{n+1})\}_{n \in \ZM_N}.
\end{equation}
Using this concrete description, the time evolution of the population of chips can be easily observed and studied in a laboratory.

Passing now to a more abstract setting, the numbers $\{N_n\}_{N\in \ZM_N}$ can be encoded in a vector from the Hilbert space $\ell^2(\ZM_N)$:
\begin{equation}
\{N_n\}_{n \in \ZM_N} \mapsto |\psi\rangle = \sum_{n \in \ZM_N} N_n |n\rangle.
\end{equation}
 The generator of the dynamics we just described can be encoded into a linear operator over same Hilbert space:
\begin{equation}\label{Eq:H0}
H = \tfrac{1}{2} \sum_{n \in \ZM_N} \big ( |n-1\rangle \langle n|+ |n-1\rangle \langle n|\big ).
\end{equation}
Indeed, one can verify that the algorithm described above translates into
\begin{equation}
|\psi\rangle \mapsto |\psi'\rangle = H |\psi\rangle.
\end{equation}
Repeating the arguments for a general discrete group $G$, one can easily see how dynamics over $\ell^2(G)$ can be generated using similar algorithms.

The type of linear operators resulting from such algorithms appear quite often in condensed matter physics, where they generate the quantum dynamics of electrons in crystals. As we already mentioned, simulating such dynamics using classical degrees of freedom is a very active area of research and various dynamical features observed or predicted for the electrons have been reproduced with mechanical, acoustic and photonic crystals. The discrete dynamics over graphs, such as the one described above, is yet another venue to reproduce interesting dynamical features observed in electronic crystals \cite{EngelhardtPRB2017,TangPRX2021}. This in fact supplies an entirely new platform for performing dynamical experiments, which could be as simple as shuffling objects between the vertices of a Cayley graph. In fact, the Cayley graph doesn't need to be accurately rendered in space. All that is needed is a proper labeling of its nodes by the group elements. This is important when dynamics over complicated patterns is investigated, such as hyperbolic and fractal crystals, which are difficult to render in our Euclidean space. Thus, the platform based on population dynamics is much more flexible then the one based on metamaterials.

Despite the apparent analogies, there is a major impediment for reproducing dynamical features seen in electronic systems with population dynamics. For example, suppose we want to visualize the eigenvectors of Hamiltonian~\eqref{Eq:H0} using the algorithm described above. One can immediately see that is impossible because these eigenvectors have complex coefficients, while the state of a pupulation always involves real positive coefficients. Furthermore, many interesting electronic models require complex coefficients while the generators of any population dynamics have real positive coefficients. This constraint, for example, prohibits us from simulating with population dynamics the celebrated Haldane model of a Chern insulator and its topological edge states \cite{HaldanePRL1988}.

The present work supplies a simple practical solution to the difficulty we just mentioned. In mathematical terms, this difficulty comes from the fact that the generators of the population dynamics over a Cayley graph of a group $G$ are all drawn from the semi-ring $\RM_+ G$, while the Hamiltonian for electrons' dynamics come in general from the complex group algebra $\CM G$. Our solution consists of a $\ZM_4$-decoration of the graph, that is, passing to the Cayley graph of $\ZM_4 \times G$ and using a natural surjective semi-ring homomorphism $\eta : \RM_+(\ZM_4 \times G) \to \CM G$. Then, any time evolution from $\CM G$ can be faithfully simulated inside the quotient $\RM_+(\ZM_4 \times G)/{\rm Ker} \, \eta$. At the practical level, this amounts to porting the complex models into models of population dynamics over the Cayley graph of $\ZM_4 \times G$ and then ``reading'' this dynamics in a precisely specified way in order to drop to the quotient space $\RM_+(\ZM_4 \times G)/{\rm Ker} \, \eta$. 

In order to keep the presentation as simple as possible, we exemplify the procedure using the simple group $G=\ZM_N$, which is use here to show how various dynamical effects can achieved and observed with population dynamics. More elaborate models, such as those simulating topological insulators, will be provided in a subsequent work. By expanding the framework from semi-rings to complex algebras, the current work will unlock general techniques coming from operator algebras, {\it e.g.} operator K-theory and index theory,  to the study of population dynamics.

\section{Population dynamics over Cayley graphs}

\begin{figure}[t]
\center
\includegraphics[width=\textwidth]{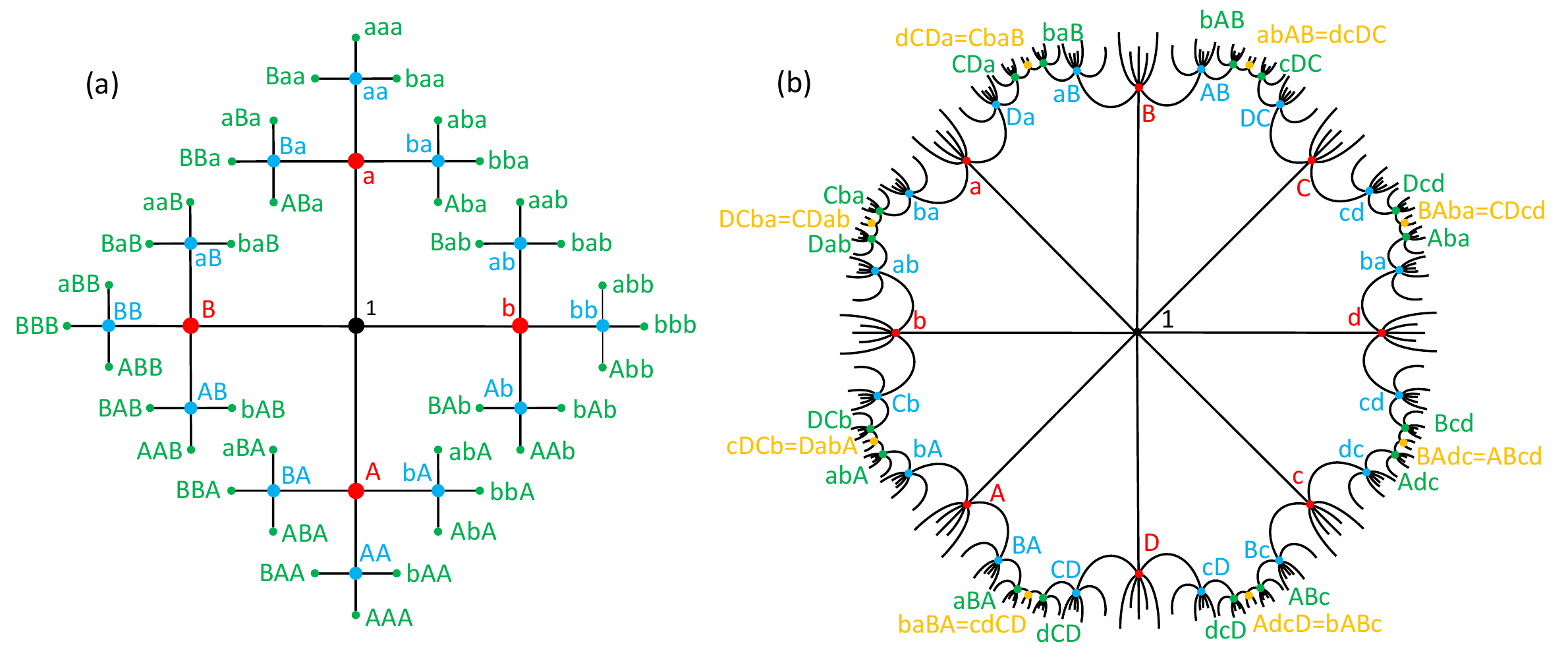}\\
  \caption{\small (a) A section of $\Cc(\FM^2(a,b),\{a,b\})$ containing all elements of length up to 3, where $A=a^{-1}$ and $B=b^{-1}$. (b) Section of $\Cc(\pi_1(\Sigma_2))$. In both panels, the color scheme is used to organize the graph by the length of the words.
}
 \label{Fig:CayleyGraphs}
\end{figure}

\subsection{Cayley graphs and diagraphs.}\label{Subsec:Cayley} Cayley graphs encode the data of a group in a geometric fashion \cite{LohBook}. For example, word problems and other theoretical problems in group theory can be  solved by inspecting these geometric objects. On the applied side, Cayley graphs can be used to generate systematic generalizations of the crystal lattices investigated in materials science. Hence they can be an abundant source of new dynamical effects, which is our main motivation for studying them.

\begin{definition} Given a discrete group $G$ and a finite subset $S$ of $G$, the Cayley graph $\Cc(G,S)$ is the un-directed graph with vertex set $G$ and edge set containing an
edge between $g$ and $sg$ whenever $g \in G$ and $s \in S$.
\end{definition}

In this definition, $S$ can any finite subset of $G$ and one should be aware that the geometry of the Cayley graph depends quite strongly on the choice of $S$. As we shall see in the explicit examples, the choice of this set $S$ is dictated by the particular models that are investigated. Many groups, however, have standard presentations in terms of generators and relations and, in such cases, there are the particular Calyley graphs that are constructed from the finite set of generators. We call them the standard Cayley graphs of the groups and denote them simply by $\Cc(G)$. Below, we supply two examples showcasing the beauty of the Cayley graphs:

\begin{figure}[t]
\center
\includegraphics[width=0.7\textwidth]{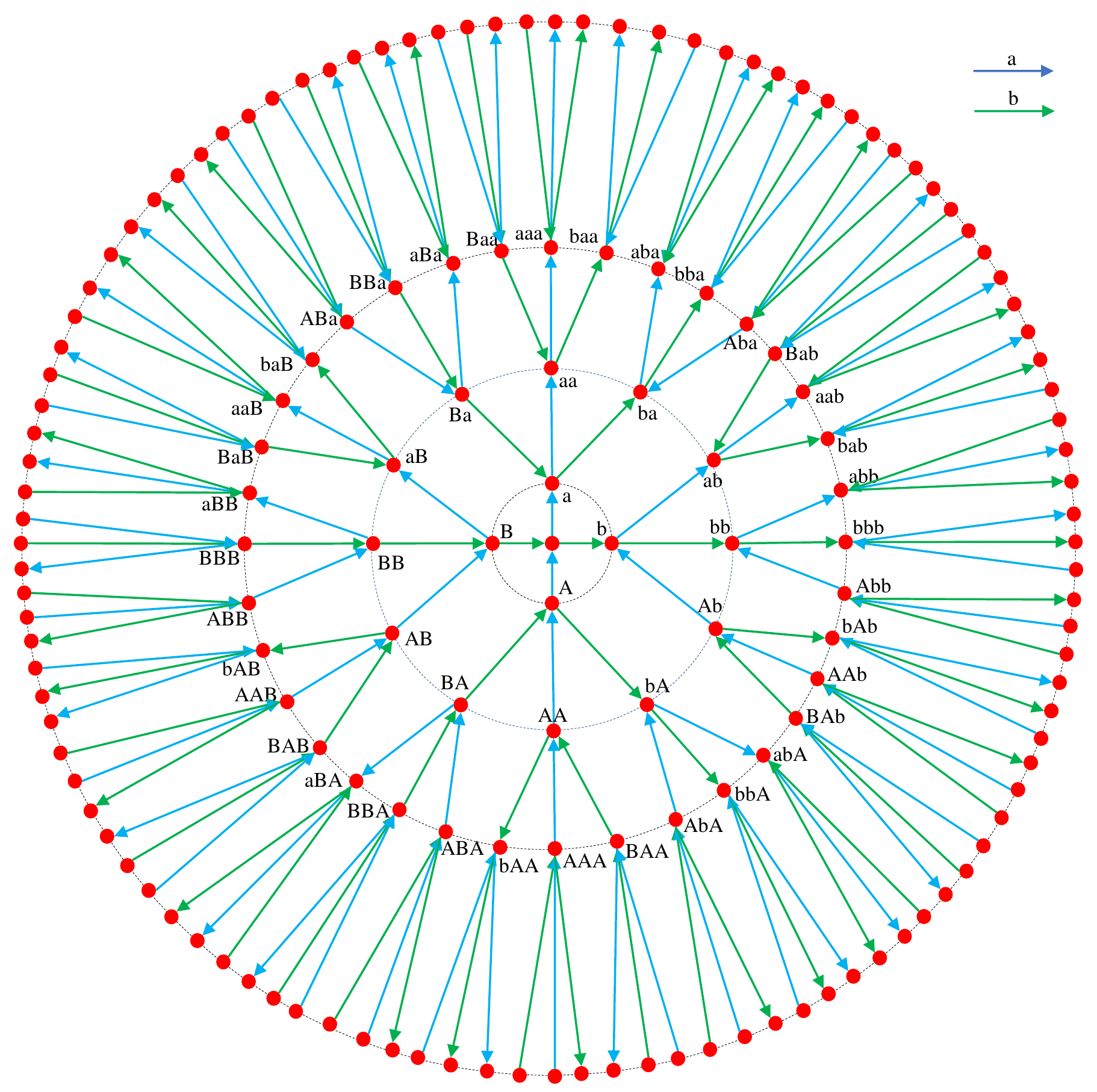}\\
  \caption{\small The Cayley diagraph of $\FM^2(a,b)$, where $A=a^{-1}$ and $B=b^{-1}$. 
}
 \label{Fig:CayleyDiGraph}
\end{figure}

\begin{example}{\rm The Cayley graph of the free abelian group with $n$ generators, $\ZM^n$, is the regular graph in the Euclidean space of dimension $n$.
}$\diamond$
\end{example}

\begin{example}{\rm The non-abelian free group $\FM^n$ with $n$-generators contains all the words made up from an alphabet of $2n$ letters, the generators and their inverses. Multiplication of two words results in the concatenation of the words. For a generic $\FM^n$ group, the Cayley  graph corresponding to the asymmetric set of generators, {\it i.e.} the standard Cayley graph, is a regular tree with coordination $n$. Such trees are referred to in the physics literature as Bethe lattices. The standard Cayley graph of $\FM^2$ is shown in Fig.~\ref{Fig:CayleyGraphs}(a). 
}$\Diamond$
\end{example}

\begin{example}{\rm The standard Cayley graph of the surface  group $\pi_1(\Sigma_2)$ is shown in Fig.~\ref{Fig:CayleyGraphs}(b), where $\Sigma_2$ is the two-hole torus. In this case, the graph displays closed cycles, which are a reflection of the non-trivial set of relations that defines this group.
}$\Diamond$
\end{example}

A more refined geometric object is the Cayley diagraph:

\begin{definition} Given a discrete group $G$ and a subset $S$ of $G$, let $c: S \to {\rm Color}$ assign a distinct color to each $s \in S$. Then the Cayley digraph $\vec \Cc(G,S,c)$ is the colored graph with vertex set $G$ and directed edges from $g$ to $sg$ for $g \in G$ and $s \in S$. All directed edges produced by $s \in S$ are assigned the color $c(s)$. 
\end{definition}

\begin{example}{\rm The standard Cayley diagraph of $\FM^2$ is shown in Fig.~\ref{Fig:CayleyDiGraph}.
}$\Diamond$
\end{example}

\subsection{Algorithms for population dynamics} Let $\Cc(G)$ be the standard Cayley graph of a discrete group $G$. We populate the nodes of this graph with identical objects, which we will call chips from now on, and denote by $N_g$ the number of chips stacked at node $g$. We are seeking algorithms that take $\{N_g\}_{g \in G}$ as input and return new values $\{N'_g\}_{g \in G}$, in a manner that respects the symmetry of the graph, in the sense that, if $\{N_g\}_{g \in G} \mapsto \{N'_g\}_{g \in G}$, then  $\{N_{gg'}\}_{g \in G} \mapsto \{N'_{gg'}\}_{g \in G}$ if the algorithm is applied to $\{N_{gg'}\}_{g \in G}$. In other words, the algorithm is invariant to the translations of the Cayley graph. Additionally, we require that the number of chips be preserved. Then, a repeated application of the algorithm will generate a discrete dynamics for the population of chips.

For this, let $S=\{g_1,\ldots,g_s\}$ be a finite sub-set of the group $G$ and consider the following protocol. The stack of $N_g$ chips sitting at site $g$ is divided in $s$ smaller stacks containing $N_g^j : = p_j N_g$ chips, $j =1,\ldots,s$, where $p_j \in [0,1]$ are fixed predefined weights that add to one: $p_1+ \cdots + p_s = 1$. After this operation is completed for each $g \in G$, the stack of $N_g^j$ chips is moved at the site $g_j g$, and this step is repeated for $g$'s and $j$'s. For finite groups, the processes we just described can be carried manually or it can be automated. Let us point out that sometimes the dynamics generated in this way can be understood more effectively if we pass from from the standard Cayley graph of $G$ to the Cayely graph generated by the set $S$. This will certainly be the case for all our applications.

At a more formal level, if the population is encoded in the vector 
\begin{equation}
|\psi \rangle = \sum_g N_g |g\rangle \in \ell^2(G),
\end{equation} 
then the dynamics we just described is given by $|\psi \rangle \mapsto D | \psi \rangle$, where
\begin{equation}\label{Eq:GenD}
D = \sum_{j=1}^s\sum_{g \in G} p_j \, |g_j  g \rangle \langle g |.
\end{equation}
Thus the operator $D$ generates the dynamics and, as such, it will be called the dynamical matrix. We point out that $D$ is a self-adjoint operator if and only if the set $S$ is invariant under taking the inverse of its elements and if $p_j = p_i$ whenever $g_i$ is the inverse of $g_j$. 

Using this formal description of the dynamics, it is easy to see that the proposed algorithms respect the two constraints we stated above. Indeed, the right action of the group $G$ on itself induces the right regular representation $\pi_R$ of $G$ on $\ell^2(G)$:
\begin{equation}
U_g |g'\rangle = |g'g^{-1} \rangle, \quad g,g' \in G.
\end{equation}
These operations are often called the translations of Cayley graph. Then one can easily check that the dynamics is invariant relative to this group action: 
\begin{equation}
\pi_R(g)D = D \pi_R(g), \quad \forall \ g \in G.
\end{equation} 
Indeed, on the basis $|g'\rangle$ of $\ell^2(G)$, we have
\begin{equation}
\pi_R(g)D |g'\rangle = \sum_{j=1}^s p_j|(g_j g') g^{-1} \rangle = \sum_{j=1}^s p_j|g_j (g' g^{-1}) \rangle = D \pi_R(g) |g'\rangle.
\end{equation} 
Thus, our protocols respect the symmetry of the Cayley graphs, a feature that is automatically present due to the associativity of the group multiplication rule. 

Furthermore, due to our constraint on the weights $\{p_j\}$, the uniform state $\psi_0$ is a right and a left eigenvector of the dynamical matrix, which is not assumed self-adjoint:
\begin{equation}
|\psi_0 \rangle : = \sum_{g \in G} |g\rangle, \quad D|\psi_0\rangle = |\psi_0\rangle, \quad D^\dagger |\psi_0\rangle = |\psi_0\rangle.
\end{equation}
In fact, it is important to note that if $\psi_0$ is a left eigenvector of $D$ then it is automatically also a right eigenvector (i.e. a left eigenvector for $D^\dagger$). Now, note that if $|\psi\rangle$ is any state of the population of chips, then 
\begin{equation}
\langle \psi_0|\psi\rangle = \sum_{g \in G} N_g,
\end{equation}
which supplies a convenient way to compute the total number of chips carried by the state $\psi$. Now, computing the number of chips for the time evolved state $D|\psi\rangle$, we get
\begin{equation}
\langle \psi_0|D|\psi\rangle = \langle D^\dagger \psi_0 | \psi \rangle = \langle \psi_0|\psi\rangle,
\end{equation}
hence the number of chips is conserved by the dynamics. An important conclusion is that $\psi_0$ being an eigenvector of $D$ with eigenvalue 1 is a simple necessary and sufficient condition for the conservation of chips during the dynamics. 

\begin{remark}{\rm In general, $N_g^j$ defined above are not natural numbers and in practice we will have to take the integer part of those values. This will result in the loss of a small number of chips at every step of the discrete dynamics. This issue will be carefully monitored during our numerical experiments.
}$\Diamond$
\end{remark}

\subsection{Group algebra: The natural environment for dynamics}
\label{Sec:GCAlg}

Given a discrete group $G$, its group algebra $\CM G$ consists of formal series
\begin{equation}\label{Eq:CG}
q = \sum_{g \in G} \alpha_g \, g, \quad \alpha_g \in \CM,
\end{equation}
where all but a finite number of terms are zero. Addition and multiplication of such formal series work in the obvious way using the group and algebraic structures of $G$ and $\CM$, respectively. In addition, there exists a natural $\ast$-operation
\begin{equation}
q^\ast = \sum_{g \in G} \bar \alpha_g \, g^{-1}, \quad (q^\ast)^\ast = q, \quad (\alpha q)^\ast = \bar \alpha q^\ast, \ \alpha \in \CM,
\end{equation}
where the bar indicates complex conjugation. Hence, $\CM G$ is naturally a $\ast$-algebra.

We denote by $e$ the neutral element of $G$. Then the map
\begin{equation}
\Tt : \CM G \to \CM, \quad \Tt(q) = \alpha_e,
\end{equation}
defines a positive trace on $\CM G$ and a pre-Hilbert structure on $\CM G$ via
\begin{equation}
\langle q, q' \rangle : = \Tt(q^\ast q'), \quad q,q' \in \CM G.
\end{equation}
The completion of the linear space $\CM G$ under this pre-Hilber structure supplies the Hilbert space $\ell^2(G)$. Indeed, one can verify that
\begin{equation}
\langle g , g' \rangle = \delta_{g,g'}, \quad g,g' \in G.
\end{equation} 
The action of $\CM G$ on itself can be extended to the action of a bounded operator on $\ell^2(G)$, and this supplies the left regular representation inside $\BM\big(\ell^2(G)\big)$, denoted by $\pi_L$ in the following. Specifically,
\begin{equation}
\pi_L(q) |g'\rangle = \sum_g \alpha_g |g g' \rangle, \quad q \in \CM G, \quad g' \in G.
\end{equation} 

Now, if $p=\sum_{j=1}^s p_j \, g_j$, then $\pi_L(p)$ is exactly the generator~\eqref{Eq:GenD} of the dynamics defined by our protocols. Also, note that $\pi_L(p^\ast) = \pi_L(p)^\dagger$ and that 
\begin{equation}
\big (\sum_{j=1}^s p_j \, g_j\big )^\ast = \sum_{j=1}^s p_j \, g_j^{-1},
\end{equation} 
from where we can quickly see when a dynamical matrix is self-adjoint or not. The main point here is that any dynamical matrix can be canonically generated from $\CM G$ and the dynamics can be studied entirely inside $\CM G$.

\begin{remark}\label{Re:IndMaps}{\rm Another reason we brought up the group algebras is because any group homomorphism $G \to H$ induces a natural homomorphism between their corresponding   group algebras. Thus, a group homomorphism creates a bridge between dynamical models generated on different Cayley graphs. Furthermore, any homomorphism between group algebras generates a natural transformation between their left regular representations. This is relevant for us because we will often jump from one group algebra to another and we don't need to explicitly specify the transformations between the respective Hilbert spaces.
}$\Diamond$
\end{remark}

\section{Quantum Dynamics Versus Population Dynamics}
\label{Sec:TheIssue}

\subsection{Main issues.} We will work with the finite abelian group $\ZM_N =\ZM/(N\, \ZM)$ and its Cayley graph. If $S$ is the generator of $\ZM_N$, then the self-adjoint operator
\begin{equation}\label{Eq:H1}
H = \tfrac{1}{2}(S + S^\ast) = \tfrac{1}{2} \sum_{n \in \ZM_N} (|n+1\rangle \langle n| + |n\rangle \langle n+1|)
\end{equation} 
can generate a quantum dynamics for, say, an electron hopping on a molecular chain, or a population dynamics via the algorithms described in the previous section. At the level of states, however, even for this simple case, there are major differences between the quantum and population dynamics. Indeed, the quantum states can have complex coefficients, while the states encoding a population must have real positive coefficients. Note that the eigenvalue problem $H \psi = \epsilon \psi$ is solved by the pairs
\begin{equation}\label{Eq:EigSys}
\Big \{ \psi_k = \Nn \sum_{n \in \ZM_N} e^{\frac{\imath 2 \pi kn}{N}} |n\rangle, \ \epsilon_k = \cos\big(\tfrac{2\pi k}{N}\big) \Big \}_{k \in \ZM_N},
\end{equation}
and, as we can see, all eigenvectors involve complex or negative coefficients, except for the uniform state $\psi_0$. The sad conclusion is that these eigenstates and their characteristics cannot be observed or demonstrated with the standard population evolution. Furthermore, for example, the following self-adjoint operator 
\begin{equation}\label{Eq:H2}
H = \tfrac{1}{4}(S +S^\ast +\imath S - \imath S).
\end{equation}
can generate a quantum dynamics, but by no means it can generate a population dynamics. Indeed, the coefficients entering the dynamical matrices must all be real and positive. Lastly, if $H$ is quantum Hamiltonian with real or complex coefficients, at first sight, there seems to impossible to simulate its quantum dynamics $e^{\imath t H} |\psi\rangle$ with population dynamics.

\subsection{The proposed solution}

Mathematically, the difference between the quantum and population dynamics over a Cayledy graph $\Cc(G)$, is that the former is associated to the group algebra $\CM G$ and its left regular representation, while the latter is associated with the semi-ring $\RM_+ G$ and its left-regular representation. Indeed, in the latter case, both the states and the Hamiltonians are constraint to take positive real values.

However, let us point out a close relation between the field of complex numbers $\CM$ and the semi-ring $\RM_+\ZM_4$, which is the backbone of our solution. For this, let $\xi$ be the generator of $\ZM_4$, hence $\ZM_4 = \{1,\xi,\xi^2,\xi^3\}$ and an element of $\RM_+\ZM_4$ accepts a unique presentation as
\begin{equation}
\sum_{j=0}^3 \beta_j \xi^j, \quad \beta_j \in \RM_+.
\end{equation}
 On the other hand, the field $\CM$ of complex numbers is also a ring, hence also a semi-ring. Then, viewing $\CM$ as a semi-ring, we note the surjective semi-ring morphism $\chi : \RM_+ \ZM_4 \to \CM$  defined by
\begin{equation}
\chi(1) = 1, \ \ \chi(\xi) =\imath, \ \ \chi(\xi^2) = -1, \ \ \chi(\xi^3) = -\imath.
\end{equation}
It has a non-trivial kernel given by the ideal
\begin{equation}
{\rm ker}\, \chi = (1+ \xi^2) \cdot \RM_+ \ZM_4.
\end{equation}
All this information can be summarized as the following exact sequence:
\begin{equation}\label{Eq:ExactSeq}
0 \to (1+ \xi^2) \cdot \RM_+ \ZM_4 \to  \RM_+ \ZM_4 \to \CM \simeq \RM_+ \ZM_4 / (1+ \xi^2) \cdot \RM_+ \ZM_4 \to 0.
\end{equation}
The conclusion is that the field $\CM$ can be generated as the quotient semi-ring
\begin{equation}
\CM = \RM_+ \ZM_4 / (1+ \xi^2) \cdot \RM_+ \ZM_4
\end{equation}
and the quotient map, of course, coincides with the map $\chi$ defined above. Furthermore, the exact sequence~\eqref{Eq:ExactSeq} is not split at the level of semi-rings, but a particularly relevant section ({\it i.e.} a right inverse for $\chi$, $\chi\circ \mathfrak s = {\rm id}$) exists at the level of linear spaces:
\begin{equation}\label{Eq:Section}
\CM \ni z=a + \imath b  \mapsto \mathfrak{s}(z)=|a| \, \xi^{1-\sgn(a)} + |b| \, \xi^{2-\sgn(b)} \in \RM_+ \ZM_4.
\end{equation} 
In fact, both $\CM$ and $\RM_+ \ZM_4$ have a structure of $\RM_+$-modules and $\mathfrak s$ is a homomorphism of such semi-modules:
\begin{equation}
\mathfrak s(r z) =r \mathfrak s(z), \quad \forall \ r \in \RM_+.
\end{equation}

Now, in order to use this simple observation to the problem of quantum and population dynamics over $\Cc(G)$, we pass to the ordinary group product  $\ZM_4 \times G$ and consider population dynamics over its Cayley graphs. In this case, the generators of the dynamics come from the group semi-ring 
\begin{equation}
\RM_+(\ZM_4 \times G) \simeq (\RM_+\ZM_4)  G.
\end{equation}
An element of this semi-ring is a formal finite sum
\begin{equation}
\tilde q = \sum_{g \in G} \tilde \alpha_g \, g, \quad \tilde \alpha_g \in \RM_+ \ZM_4.
\end{equation}
Note that the commutative semi-ring $\RM_+ \ZM_4$ is canonically embedded in $(\RM_+\ZM_4)  G$ via
\begin{equation}
\RM_+ \ZM_4 \ni \tilde \alpha \mapsto \tilde \alpha \, e \in (\RM_+\ZM_4)  G.
\end{equation}
Obviously $(\tilde \alpha \, e)\tilde q = \tilde q (\tilde \alpha \, e)$, hence $(\RM_+\ZM_4)  G$ has a structure of semi-algebra over the commutative ring $\RM_+\ZM_4$.

The essential observation is that $(1+ \xi^2) \cdot (\RM_+ \ZM_4)G$ is an ideal of the above semi-ring and that
\begin{equation}
(\RM_+\ZM_4) G/(1+ \xi^2) \cdot (\RM_+\ZM_4) G \simeq \CM G.
\end{equation}
The quotient map from $(\RM_+\ZM_4) G$ to $\CM G$ is implemented by a straightforward extension of the map $\chi$ (denote by the same symbol $\chi$):
\begin{equation}\label{Eq:Chi}
(\RM_+ \ZM_4) G \ni \sum_{g \in G} \tilde \alpha_g \, g \mapsto \sum_{g \in G} \chi(\tilde \alpha_g) \, g \in \CM G.
\end{equation}
It is important to keep in mind that this map, denoted by the same symbol $\chi$, is a semi-ring homomorphism. Thus, it respects the addition and multiplication operations and, as such, it commutes with the functional calculus, in the sense that
\begin{equation}
\chi\big (\phi(\tilde q)\big ) = \phi\big (\chi(\tilde q))
\end{equation}
for any polynomial $\phi$. Lastly, the section defined in Eq.~\eqref{Eq:Section} extends to a right inverse for $\eta$,
\begin{equation}\label{Eq:Sec2}
\CM G \ni \sum_{g \in G} \alpha_g \, g \mapsto \sum_{g \in G} \mathfrak{s}(\alpha_g) \, g \in (\RM_+\ZM_4) G.
\end{equation}
This extension will be denoted by the same symbol $\mathfrak s$ and note that $\mathfrak s$ is a homomorphism between $\RM_+$-modules.

The map $\chi$ and its right inverse $\mathfrak s$ are our essential tools that will enable us to port models from $\CM G$ into models from $\RM_+(\ZM_4 \times G)$ and, as such, to simulate a quantum dynamics on $\Cc(G)$ with a population dynamics over $\Cc(\ZM_4 \times G)$. Specific experiments enabled by our solution will are discussed next.

\subsection{Applications}\label{Subsec:Applications} One interesting experiment is visualization of stationary quantum states. Specifically, if $H$ is a quantum Hamiltonian and $\psi_\epsilon$ is an eigenvector $H \psi_\epsilon = \epsilon \psi_\epsilon$, we can generate a dynamical matrix and an initial state for a population dynamics on $\Cc(\ZM_4 \times G)$ as
\begin{equation}
\tilde D = \mathfrak s(H), \quad |\tilde \psi_\epsilon \rangle = |\mathfrak s(\psi_\epsilon)\rangle,
\end{equation}
where for $\tilde \psi$ we used Remark~\ref{Re:IndMaps}.
We can then use our algorithm to time-evolve the population
\begin{equation}
|\tilde \psi_\epsilon(t) \rangle : = \tilde D^t |\tilde \psi_\epsilon \rangle, \quad t \in \NM.
\end{equation}
Projecting back on $\CM G$ and its left regular representation, we will find that
\begin{equation}
\big|\chi\big (\tilde \psi_\epsilon(t) \big )\big\rangle = \chi\Big (\tilde D^t |\tilde \psi_\epsilon \rangle \Big )=\chi \big(\tilde D^t \big )|\chi(\tilde \psi_\epsilon) \rangle = \chi \big(\tilde D \big )^t|\psi_\epsilon \rangle = H^t |\psi_\epsilon \rangle,
\end{equation}
hence, apart from a scaling factor, this projected state is stationary under the discrete time evolution. As one can see, the property of the map $\chi$ of respecting the addition and multiplication is essential for our arguments.

To be effective, we must supply an experimental protocol for applying the quotient map~\eqref{Eq:Chi}. For this, an experimenter examines a coefficient $\tilde \alpha_g = \sum_{j=0}^3 \alpha_g^j \xi^j$, which is encoded in the four stacks $\alpha_g^j$ of chips. Now, 
\begin{equation}
\chi(\tilde \alpha_g) = \alpha_g^0-\alpha_g^3 + \imath(\alpha_g^1 - \alpha_g^4),
\end{equation} 
thus, the experimenter simply needs to compare the stacks $\alpha_g^0$ and $\alpha_g^3$ of chips, remove a number of chips from a stack such that the two stacks remain with equal number of chips, and place those chips in a stack called $\alpha_g^r$ above the zero level if $\alpha_g^0>\alpha_g^3$ and below the zero level otherwise. Similarly from the stacks $\alpha_g^1$ and $\alpha_g^4$, the experimenter creates the stack $\alpha_g^i$. For a state 
\begin{equation}
|\tilde \psi\rangle = \sum_{g \in G} \tilde \alpha_g |g\rangle = \sum_{j=0}^3\sum_{g \in G} \alpha_g^j |\xi^j,g\rangle \in \ell^2(\ZM_4 \times G),
\end{equation}
by repeating the procedure for all $\tilde \alpha_g$'s, the experimenter creates the stacks $\{\alpha_g^r,\alpha_g^i\}$ encoding the real and imaginary parts of $|\chi(\tilde \psi)\rangle$. This gives a simple protocol for visualizing quantum states with a population of chips.

Another class of quantum experiments that can be simulated with a population of chips is visualizing the time evolution of a quantum state
\begin{equation}
|\psi(t) \rangle = e^{\imath t H} |\psi\rangle, \quad t \in \RM_+.
\end{equation}
For this, we use the standard approximation of the exponential function
\begin{equation}
e^{\imath t H} = \lim_{m \to \infty} \big ( 1+ \tfrac{\imath t}{m}H \big )^m
\end{equation}
and generate a dynamical matrix and an initial population of chips
\begin{equation}\label{Eq:DTEv}
\tilde D = \gamma \, \mathfrak s\big(1 +  \tfrac{\imath t}{m}H \big ), \quad |\tilde \psi(0)\rangle = |\mathfrak s (\psi)\rangle,
\end{equation}
where $\gamma$ is a positive factor that ensures that the number of chips are conserved. We then perform the discrete time evolution
\begin{equation}
|\tilde \psi(k) \rangle  = \tilde D |\tilde \psi(k-1)\rangle, \quad k=1,\ldots,m,
\end{equation}
on the Cayley graph of $\ZM_4 \times G$ and finally project back on $\CM G$ and its left regular representation, to find:
\begin{equation}
\gamma^{-m} \, \big|\chi\big (\tilde \psi(m) \big )\big\rangle = \gamma^{-m} \, \chi \big(\tilde D \big )^m|\chi(\psi) \rangle = \big ( 1+ \tfrac{\imath t}{m}H \big )^m |\psi \rangle \approx e^{\imath t H}|\psi\rangle.
\end{equation}
Concrete experiments along these lines are presented in the next section.

\section{Examples}
\label{Sec:Example1}

\begin{figure}[t]
\center
\includegraphics[width=\textwidth]{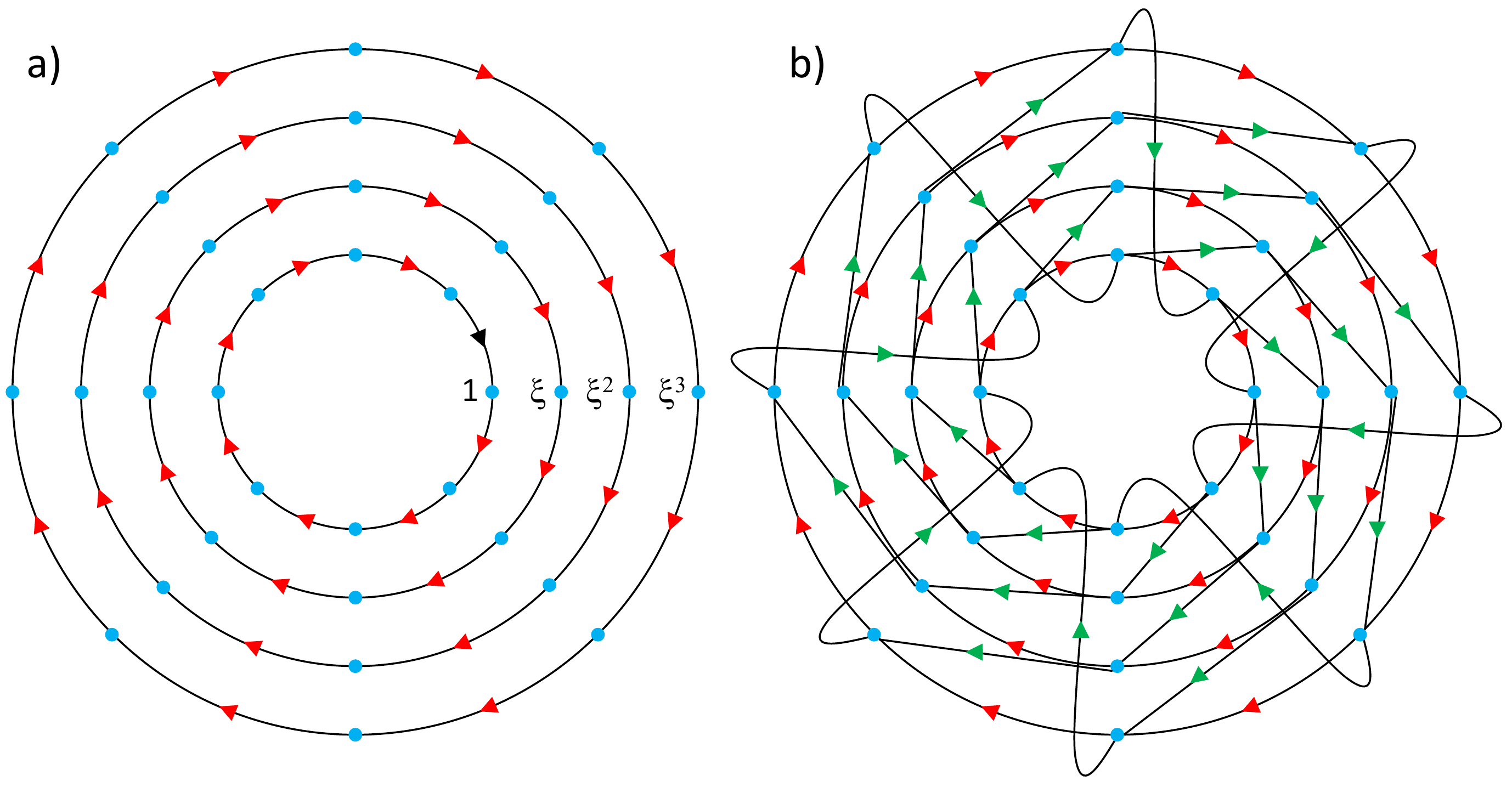}\\
  \caption{\small Cayley diagraph corresponding to: a) $(\ZM_4 \otimes \ZM_N), \{S\} )$, and b)  $(\ZM_4 \otimes \ZM_N), \{S,\xi S\} )$, for $N=8$. Here, the red arrows correspond to the action of $S$, the generator of $\ZM_N$, and the green arrows to the element $\xi S$.
}
 \label{Fig:Cayley12}
\end{figure}

\subsection{Experiment 1} In this experiment, we will demonstrate and visualize a stationary quantum state. For this, we consider the quantum Hamiltonian $H=\frac{1}{2}(S+S^\ast)$ introduced in Sec.~\ref{Sec:TheIssue}, which now is viewed as an element of $\RM_+(\ZM_4 \times \ZM_N)$. To convey this fact, we will work with $\tilde D = \mathfrak(H)$, even though $\tilde D$ has the same expression as $H$. Fig.~\ref{Fig:Cayley12} shows the Cayley diagraph corresponding to the group $\ZM_4 \times \ZM_N$ and element $S \in \ZM_N$. We call this the $\ZM_4$-decorated Cayley graph. For the Hamiltonian~\eqref{Eq:H1}, the decorated dynamics remains very simple because this Hamiltonian has real positive coefficients. 

\begin{figure}[t]
\center
\includegraphics[width=\textwidth]{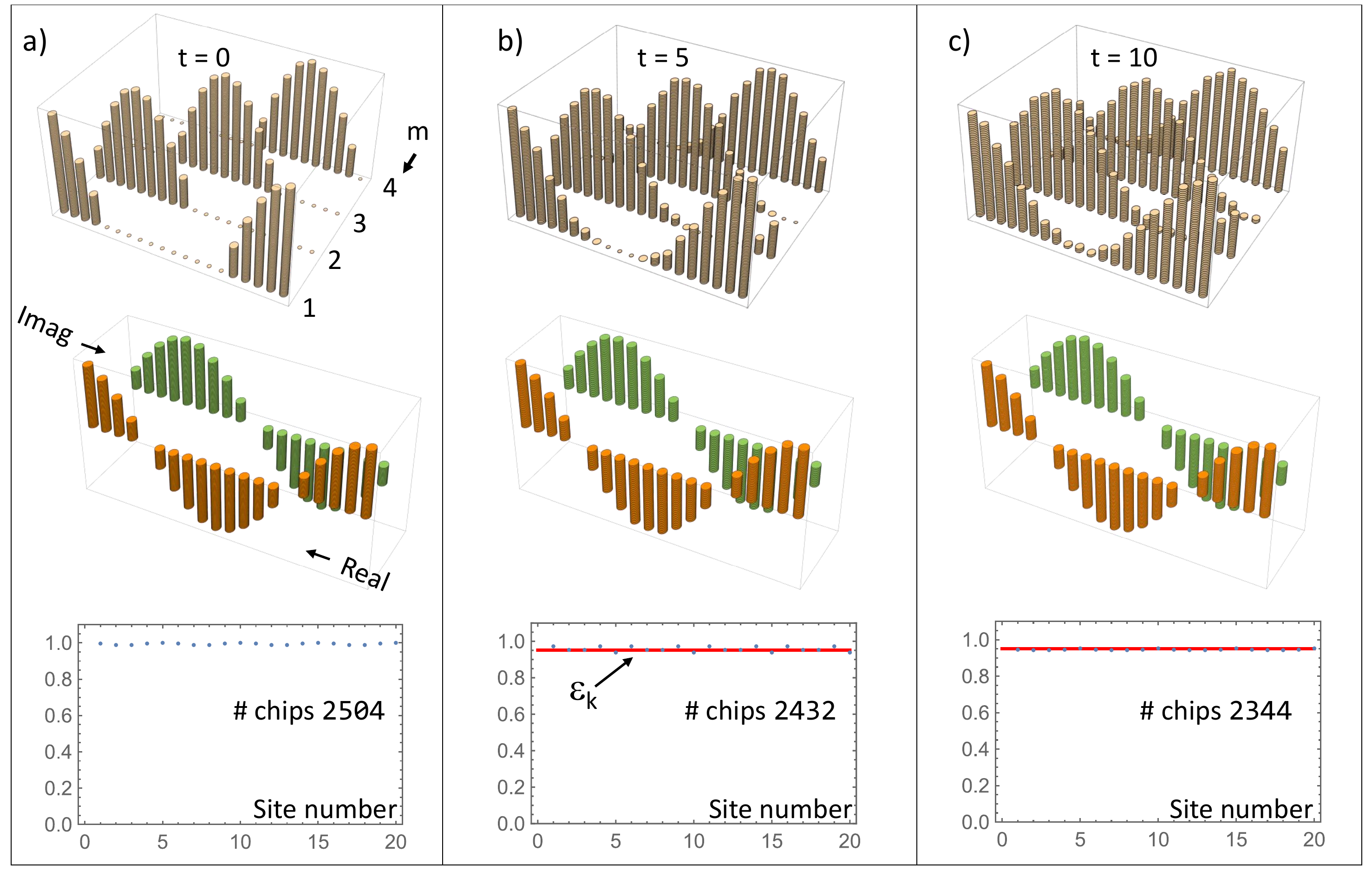}\\
  \caption{\small $\ZM_4$-decorated time evolution $\psi(t) = H^t \psi(t)$ for a state prepared as $\psi(0)=\mathfrak s(\psi_k)$, with $\psi_k$ as in Eq.~\eqref{Eq:EigSys}, together with appropriate tests, for $k=1$ and  the indicated values of $t$. The graphs at the bottom show the absolute values of the ratios of the coefficients of $\eta(\psi(0))$ and $\psi_k$ (panel b) and of the the absolute values of the ratios of the coefficients of $\eta(\psi(t))$ and $\eta(\psi(t-1))$, as compared to $\epsilon_k$ from Eq.~\eqref{Eq:EigSys} and indicated by the red line (panels d and f). The analysis was generated with $N=20$.
}
 \label{Fig:Example1}
\end{figure}

\begin{figure}[t]
\center
\includegraphics[width=\textwidth]{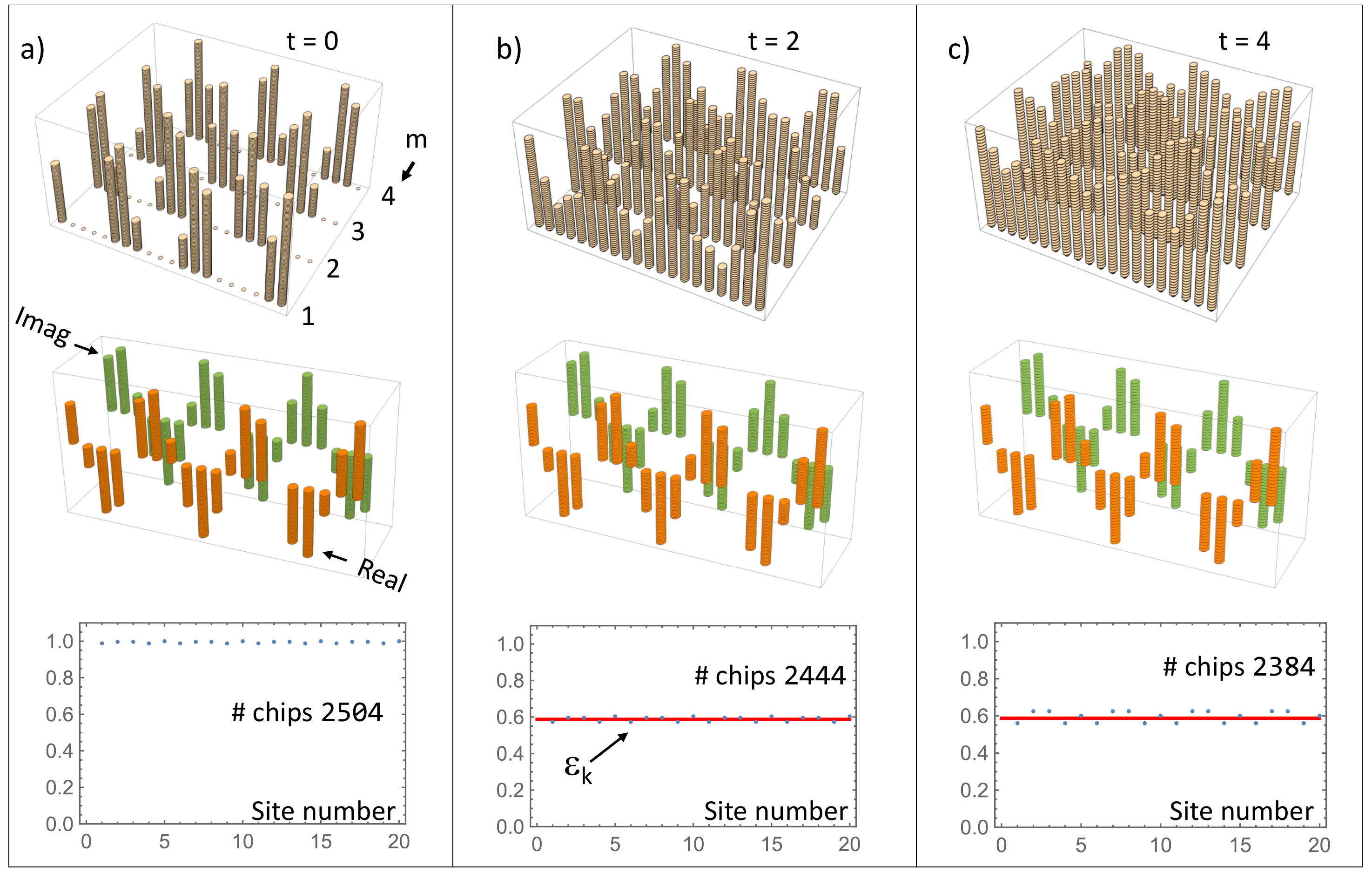}\\
  \caption{\small Same as Fig.~\ref{Fig:Example1}, but for $k=3$.
}
 \label{Fig:Example2}
\end{figure}

Now, we choose a state $\psi_k$ as in Eq.~\eqref{Eq:EigSys} and ported it on $\ell^2(\ZM_4 \times \ZM_N)$ using the section $\mathfrak s$ from Eq.~\eqref{Eq:Sec2}:
\begin{equation}
\begin{aligned}
\tilde \psi_k = \mathfrak{s} (\psi_k) = \Nn \sum_{n \in \ZM_N} & \big [|\cos\big(\tfrac{2 \pi kn}{N}\big)| \big | \xi^{1-\sgn(\cos(\frac{2 \pi kn}{N}))},n\big \rangle  \\
& \qquad \qquad +|\sin\big(\tfrac{2 \pi kn}{N}\big)| \big | \xi^{1-\sgn(\sin(\frac{2 \pi kn}{N}))},n \big \rangle \big ].
\end{aligned}
\end{equation} 
The population corresponding to this state is reported in the top row of Fig.~\ref{Fig:Example1}(a), together with the visualization of $\chi(\tilde \psi_k)$ (second row) and a comparison between $\chi(\tilde \psi_k)$ and $\psi_k$ (third row), for $k=1$. The data seen in this figure confirms that the maps $\chi$ and $\mathfrak s$ work as expected. Same information is reported in Fig.~\ref{Fig:Example2}(a) for the stationary state generated with $k=3$. Panels (b) and (c) in Fig.~\ref{Fig:Example1}(a) report the states $|\tilde \psi(t)\rangle = \tilde D^t |\tilde \psi\rangle$ for $t=5$ and $t=10$, respectively, visualizations of $|\psi(t)\rangle = \big|\chi\big (\tilde \psi(t)\big ) \big \rangle$, together with checks of their expected relations $|\psi(t)\rangle = \epsilon_k |\psi(t-1)\rangle$. Same information is reported in Fig.~\ref{Fig:Example2}(b,c) for the case $k=3$ and times $t=2$ and $t=4$, respectively.

These experiments reveal several interesting observations. First, the time-evolved population dynamics on the $\ZM_4$-decorated graph produces somewhat complicated states, yet the projection onto the original graph, using the protocol described in the previous section, produces clear stationary states. As we already explained, there is a loss of chips during the dynamics, but this obviously does not affect the demonstrations we actually want to show.

\subsection{Experiment 2.} Here we consider the following Hamiltonian on $\ell^2(\ZM_N)$:
\begin{equation}\label{Eq:H2}
H = \tfrac{1}{4}(S +S^\ast +\imath S - \imath S).
\end{equation}
It has the same eigenvectors $\psi_k$ but with corresponding eigenvalues
\begin{equation}
\epsilon_k = \tfrac{1}{2}\big [ \cos\big ( \tfrac{2\pi k}{N} \big ) + \sin \big ( \tfrac{2\pi k}{N} \big )\big ].
\end{equation}
Note that these eigenvalues can be positive and negative as well. Having complex coefficients, the dynamics of this Hamiltonian cannot be directly associated with a population dynamics. We show, however, that its dynamics can be studied and observed with population dynamics experiments, if we use our strategy and port $H$ on $\ell^2 (\ZM_4 \times \ZM_N )$ as:
\begin{equation}\label{Eq:EtaH2}
H \mapsto \tilde D = \tfrac{1}{4}(S +S^\ast +\xi S + \xi^3 S).
\end{equation}

The dynamics of generated by $\tilde D$ can be understood from the Cayley diagraph corresponding to the elements involved in $\tilde D$, namely, $S$ and $\xi S$. This Cayley diagraph is shown in Fig.~\ref{Fig:Cayley12}(b). Note that $\xi^3 S^\ast =(\xi S)^{-1}$, so this element is already covered be the shown Cayley diagraph. 

\begin{figure}[t]
\center
\includegraphics[width=\textwidth]{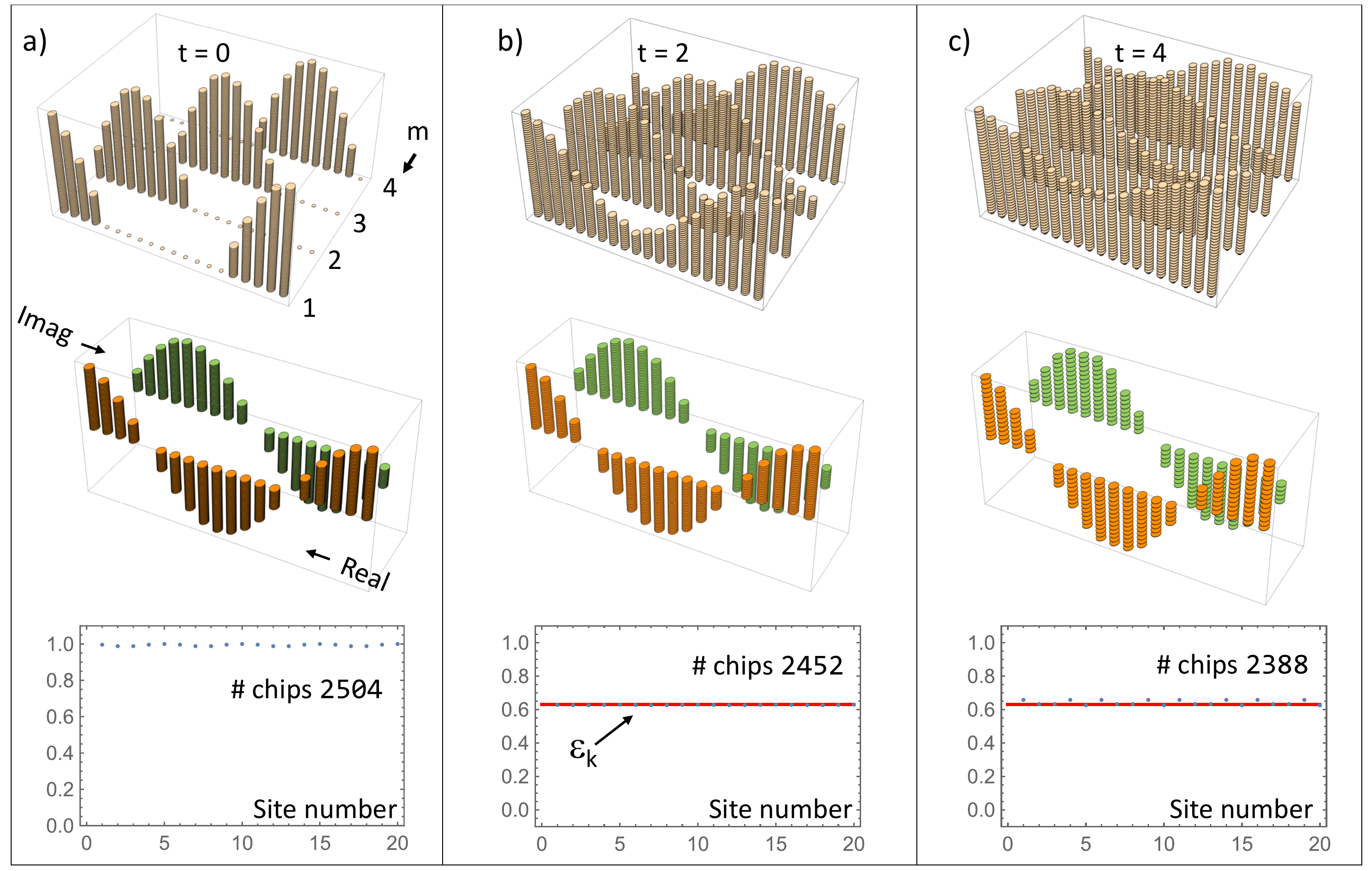}\\
  \caption{\small Same as Fig.~\ref{Fig:Example1}, but for Hamiltonian~\eqref{Eq:H2}.
}
 \label{Fig:Example3}
\end{figure}

To demonstrate and visualize the stationary states of $H$, we map $\psi_k$ on the Cayley graph of of $\ZM_4 \times \ZM_n$ and act with the dynamical matrix $\tilde D$, as it was done in the previous case. This dynamics is reported in the top row of Fig.~\ref{Fig:Example3} for the state generated with $k=1$. A projection of $\tilde \psi_k(t)$ back on the left regular representation of $\CM G$ reveals again the stationary character of the quantum state $\psi_k$.

\subsection{Experiment 3.} In this experiment, we reproduce the quantum dynamics $|\psi(t)\rangle = e^{\imath t H}|\psi(0)\rangle$, with $H$ as in Eq.~\eqref{Eq:H1} and an initial state localized entirely at one site, $\langle n |\psi(0)\rangle = \delta_{n,N/2}$. For this, we follow literally the procedure detailed in subsection~\ref{Subsec:Applications} and we calculate the dynamical matrix $\tilde D$ introduced in Eq.~\eqref{Eq:DTEv}, which takes the specific form
\begin{equation}\label{Eq:D10}
\tilde D = \gamma \big(1 +\tfrac{t}{m}(\xi S +\xi S^\ast)\big ), \quad \gamma = 1/(1+2t/m).
\end{equation}
Note that this dynamical matrix is not self-adjoint but it conserves the number of chips. The initial vector on the decorated Cayley graph is simply 
\begin{equation}\label{Eq:Psi0}
\langle j,n|\tilde \psi(0)\rangle =\delta_{j,1} \delta_{n,N/2}, \quad j=\overline{1,4},\quad  n \in \ZM_N.
\end{equation} 
To implement the population dynamics, however, we will place a large number of chips at the location $(1,N/2)$ to ensure that the shuffling doesn't stop due to lack of chips. This is not an isuue because, afterwards, the time-evolved states are properly normalized, hence taking out the arbitrariness introduced by this little detail.

\begin{figure}[t]
\center
\includegraphics[width=\textwidth]{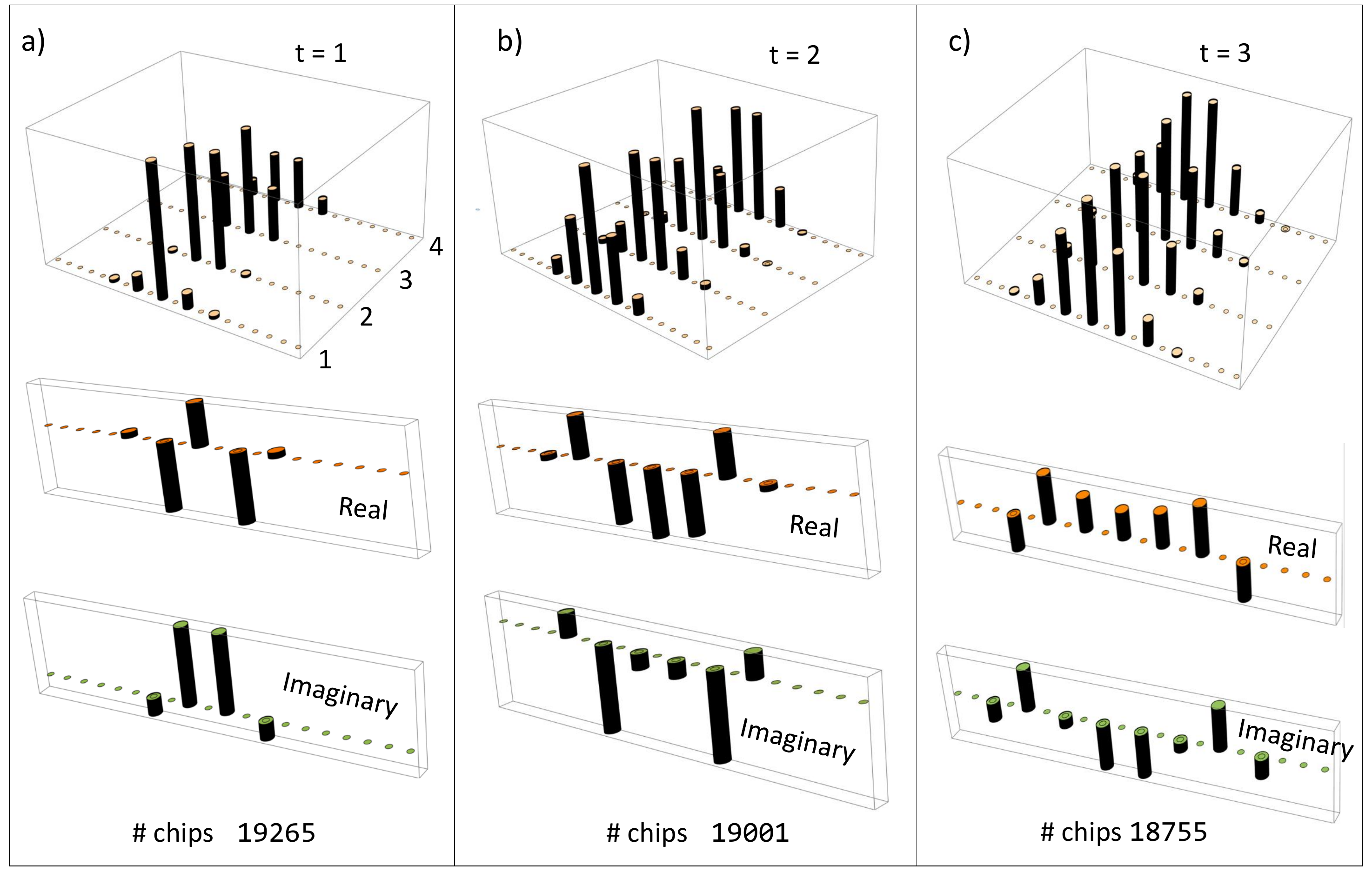}\\
  \caption{\small Quantum evolution simulated with population dynamics over the decorated Cayley graph $\Cc(\ZM_4 \times \ZM_N)$. The top rows show the populations of chips corresponding to the states $|\tilde \psi(t)\rangle=\tilde D^m |\tilde \psi(0)\rangle$, with $\tilde D$ as in Eq.~\eqref{Eq:D10} and $|\tilde \psi(0)\rangle$ as in Eq.~\eqref{Eq:Psi0}. The second and third rows show the real and imaginary parts of the projections $\big |\chi\big(\tilde \psi(t)\big)\big \rangle$. The numerical experiments were carried out with $N=20$, $m=100$ and the initial number of chips was $20,000$.
}
 \label{Fig:Example4}
\end{figure}

The top row of Fig.~\ref{Fig:Example4} shows the time evolution $|\tilde \psi(t)\rangle$ of the initial state $|\tilde \psi(0)\rangle$ under the dynamical matrix $\tilde D$ from Eq.~\eqref{Eq:D10}. The second and third rows of the same figure show the real and imaginary parts of the projections $\big|\chi\big(\tilde \psi(t)\big)\big \rangle$ for different times. The normalizations of these population states coincide with the exact time-evolved quantum states $|\psi(t)\rangle$, up to less than 1\% differences.

\end{document}